\documentclass{article}
\usepackage{amssymb}

\usepackage{makeidx}
\usepackage{graphicx}
\usepackage{amsmath}

\input{tcilatex}

\begin{document}

\author{S. Manoff$\thanks{%
Permanent address: Bulgarian Academy of Sciences, Instutute for Nuclear
Research and Nuclear Energy, Department of Theoretical Physics, Blvd.
Tzarigradsko Chaussee 72, 1784 Sofia - Bulgaria}$, B. Dimitrov \\
\textit{Joint Institute for Nuclear Research,}\\
\textit{Bogoliubov Laboratory of Theoretical Physics,}\\
\textit{Dubna, Moscow Region}\\
\textit{141980 Russia}}
\title{Weyl's spaces with shear-free and expansion-free conformal Killing vectors
and the motion of a free spinless test particle.}
\date{e-mail addresses: \textit{smanov@thsun1.jinr.ru, smanov@inrne.bas.bg,
bogdan@thsun1.jinr.ru}}
\maketitle

\begin{abstract}
Conditions for the existence of shear-free and expansion-free non-null
vector fields in spaces with affine connections and metrics are found. On
their basis Weyl's spaces with shear-free and expansion-free conformal
Killing vectors \ are considered. The necessary and sufficient conditions
are found under which a free spinless test particle could move in spaces
with affine connections and metrics on a curve described by means of an
auto-parallel equation. In Weyl's spaces with Weyl's covector, constructed
by the use of a dilaton field, the dilaton field appears as a scaling factor
for the rest mass density of the test particle.

\end{abstract}

\section{Introduction}

In the last years Weyl's spaces have deserved some interest related to the
possibility of using mathematical models of space-time different from
(pseudo) Riemannian spaces without torsion ($V_{n}$-spaces) or with torsion (%
$U_{n}$-spaces)\cite{Hehl-1} $\div $ \cite{Salim}. On the one side, Weyl's
spaces appear as a generalization of $V_{n}$- and $U_{n}$-spaces. On the
other side, they are special cases of spaces with affine connections and
metrics. The use of spaces with affine connections and metrics as models of
space-time has been critically evaluated from different points of view \cite
{Hayashi}, \cite{Treder}. But recently, it has been proved that in spaces
with contravariant and covariant affine connections (whose components differ
only by sign) and metrics [$(L_{n},g)$-spaces] as well as in spaces with
contravariant and covariant affine connections (whose components differ not
only by sign) and metrics [$(\overline{L}_{n},g)$-spaces] \cite{Manoff-1}, 
\cite{Manoff-1a} the principle of equivalence holds \cite{Iliev-1}$\div $%
\cite{Hartley}. In these spaces special types of transports (called
Fermi-Walker transports) \cite{Manoff-2} $\div $ \cite{Manoff-3} exist which
do not deform a Lorentz basis. Therefore, the low of causality is not abused
in $(L_{n},g)$- and $(\overline{L}_{n},g)$-spaces if one uses a Fermi-Walker
transport instead of a parallel transport (used in a $V_{n}$-space).
Moreover, there also exist other types of transports (called conformal
transports) \cite{Manoff-4}, \cite{Manoff-5} under which a light cone does
not deform. At the same time, the auto-parallel equation can play the same
role in $(L_{n},g)$- and $(\overline{L}_{n},g)$-spaces as the geodesic
equation does in the Einstein theory of gravitation (ETG) in $V_{n}$-spaces 
\cite{Manoff-6}, \cite{Manoff-7}. On this basis, many of the
differential-geometric construction used in the ETG in $V_{4}$-spaces could
be generalized for the cases of $(L_{n},g)$- and $(\overline{L}_{n},g)$%
-spaces, and especially for Weyl's spaces without torsion ($W_{n}$ or $%
\overline{W}_{n}$-spaces) or in Weyl's spaces with torsion ($Y_{n}$- or $%
\overline{Y}_{n}$-spaces) as special cases of $(L_{n},g)$- or $(\overline{L}%
_{n},g)$-spaces. Bearing in mind this background a question arises about
possible physical applications and interpretation of mathematical
constructions from ETG generalized for Weyl's spaces. It is well known that
every classical field theory over spaces with affine connections and metrics
could be considered as a theory of continuous media in these spaces \cite
{Hehl-3} $\div $ \cite{Manoff-8b}. On this ground, notions of the continuous
media mechanics (such as deformation velocity and acceleration, shear
velocity and acceleration, rotation velocity and acceleration, expansion
velocity and acceleration) have been used as invariant characteristics for
spaces admitting vector fields with special kinematic characteristics \cite
{Ehlers}, \cite{Stephani}.

\subsection{Problems and results}

The main task of this paper \ is the investigation of Weyl's spaces with
respect to their ability to admit conformal contravariant Killing vector
fields that are shear-free and expansion-free vector fields. On this basis,
conditions for the existence of shear-free and expansion-free non-null
(non-isotropic) vector fields in spaces with affine connections and metrics
are found and then specialized for Weyl's spaces. At the same time, a
possible interpretation of a dilaton field, appearing in the structure of
special types of Weyl's spaces, is found on the basis of the auto-parallel
equation describing the motion of a free spinless test particle in these
types of spaces.

In Section 2 some recurrent relations in spaces with affine connections and
metrics are considered. The equivalence of the action of the Lie
differential operator and of the covariant differential operator on the
invariant volume element is introduced as condition for the metrics and for
a vector field along which these operators act. In Section 3 some properties
of Weyl's spaces are recalled related to the properties of scalar and tensor
fields over such type of spaces. In Section 4 the notion of the relative
velocity and its kinematic characteristics in spaces with affine connections
and metrics related to the sear, rotation, and expansion velocities are
introduced. It is shown that the equivalence condition, proposed in Section
2, appears as a condition for the existence of shear-free and expansion-free
non-null vector fields in spaces with affine connections and metrics. The
same condition in a Weyl's space appears as a condition for the existence of
a shear-free and expansion-free conformal Killing vector field of special
type. Sufficient conditions are found under which in special type of Weyl's
spaces non-null auto-parallel, shear-free, and expansion-free conformal
Killing vector fields could exist. In Section 5 the auto-parallel equation
in Weyl's spaces is discussed as an equation for describing the motion of a
free moving spinless test particle. Concluding remarks comprise the final
Section 6. The most considerations are given in details (even in full
details) for those readers who are not familiar with the investigated
problems.

\subsection{Abbreviations, definitions, and symbols}

In the further considerations in this paper we will use the following
abbreviations, definitions and symbols:

:= means by definition.

$M$ is a symbol for a differentiable manifold with $\dim M=n$. $T(M):=\cup
_{x\in M}T_{x}(M)$ and $T^{\ast }(M):=\cup _{x\in M}T_{x}^{\ast }(M)$ are
the tangent and the cotangent spaces at $M$ respectively.

$(\overline{L}_{n},g)$, $\overline{Y}_{n}$, $\overline{U}_{n}$, and $%
\overline{V}_{n}$ are spaces with contravariant and covariant affine
connections and metrics whose components \textit{differ not only by sign } 
\cite{Manoff-1}. In such type of spaces the non-canonical contraction
operator $S$ acts on a contravariant basic vector field $e_{j}$ (or $%
\partial _{j}$)$\,\in \{e_{j}$ (or $\partial _{j}$)$\}\subset T(M)$ and on a
covariant \ basic vector field $e^{i}\,$\ (or $dx^{i}$) $\in \{e^{i}$ (or $%
dx^{i}$)$\}\subset T^{\ast }(M)$ in the form 
\begin{eqnarray*}
S &:&(e^{i},e_{j})\longrightarrow S(e^{i},e_{j}):=S(e_{j},e^{i}):=f^{i}\,_{j}%
\text{ , \ \ \ \ \ \ } \\
\text{\ \ \ \ }f^{i}\,_{j} &\in &C^{r}(M)\text{ , \ \ \ \ \ \ \ \ }r\geqq 2%
\text{ , \ \ \ \ \ \ }\det (f^{i}\,_{j})\neq 0\text{, \ \ \ \ \ \ } \\
\exists \text{\ \ \ }f_{i}\,^{k} &\in &C^{r}(M)\text{, \ \ \ \ \ \ \ \ \ }%
r\geqq 2\text{ }:\text{\ \ \ \ \ \ \ \ }f^{i}\,_{j}\cdot
f_{i}\,^{k}:=g_{j}^{k}\text{, \ \ \ \ \ \ \ \ \ \ }
\end{eqnarray*}

In these spaces, for example, $g(u)=g_{ik}\cdot f^{k}\,_{j}\cdot u^{j}\cdot
dx^{i}:=g_{i\overline{j}}\cdot u^{j}\cdot dx^{i}=g_{ij}\cdot u^{\overline{j}%
}\cdot dx^{i}:=u_{i}\cdot dx^{i}$, $g(u,u)=g_{kl}\cdot f^{k}\,_{i}\cdot
f^{l}\,_{j}\cdot u^{i}\cdot u^{j}:=g_{\overline{i}\overline{j}}\cdot
u^{i}\cdot u^{j}=g_{ij}\cdot u^{\overline{i}}\cdot u^{\overline{j}%
}=u_{j}\cdot u^{\overline{j}}:=u_{\overline{i}}\cdot u^{i}$, $g^{\overline{i}%
\overline{j}}\cdot g_{jk}=\delta _{k}^{i}=g_{k}^{i}$, $g_{\overline{i}%
\overline{k}}.g^{kj}=g_{i}^{j}$. The components $\delta _{j}^{i}:=g_{j}^{i}$
($\mid =0$ for $i\neq j$ and $\mid =1$ for $i=j$) are the components of the
Kronecker tensor $Kr:=g_{j}^{i}\cdot \partial _{i}\otimes dx^{j}$.

$(L_{n},g)$, $Y_{n}$, $U_{n}$, and $V_{n}$ are spaces with contravariant and
covariant affine connections and metrics whose components \textit{differ
only by sign} \cite{Norden}, \cite{Raschewski}. In such type of spaces the
canonical contraction operator $S:=C$ acts on a contravariant basic vector
field $e_{j}$ (or $\partial _{j}$)$\,\in \{e_{j}$ (or $\partial _{j}$)$%
\}\subset T(M)$ and on a covariant \ basic vector field $e^{i}\,$\ (or $%
dx^{i}$) $\in \{e^{i}$ (or $dx^{i}$)$\}\subset T^{\ast }(M)$ in the form 
\begin{equation*}
C:(e^{i},e_{j})\longrightarrow C(e^{i},e_{j}):=C(e_{j},e^{i}):=\delta
_{j}^{i}:=g_{j}^{i}\text{ .}
\end{equation*}

In these spaces, for example, $g(u)=g_{ik}\cdot g_{j}^{k}\cdot u^{j}\cdot
dx^{i}:=g_{ij}\cdot u^{j}\cdot dx^{i}=u_{i}\cdot dx^{i}$, $%
g(u,u)=g_{kl}\cdot g_{i}^{k}\cdot g_{j}^{l}\cdot u^{i}\cdot
u^{j}:=g_{ij}\cdot u^{i}\cdot u^{j}=u_{i}\cdot u^{i}$.

\textit{Remark}. All results found for $(\overline{L}_{n},g)$-spaces could
be specialized for $(L_{n},g)$-spaces by omitting the bars above or under
the indices.

$\nabla _{u}$ is the covariant differential operator acting on the elements
of the tensor algebra $\mathcal{T}$ over $M$. The action of \ $\nabla _{u}$
is called covariant differentiation (covariant transport) along a
contravariant vector field $u$, for instance, 
\begin{equation}
\nabla _{u}v:=v_{\;;j}^{i}\cdot u^{j}\cdot \partial
_{i}=(v^{i}\,_{,j}+\Gamma _{kj}^{i}\cdot v^{k})\cdot u^{j}\cdot \partial _{i}%
\text{ , \ \ \ \ \ }v\in T(M)\text{ ,}  \label{0.1}
\end{equation}
where $v^{i}\,_{,j}:=\partial v^{i}/\partial x^{j}$ and $\Gamma _{jk}^{i}$
are the components of the contravariant affine connection $\Gamma $ in a
contravariant co-ordinate basis $\{\partial _{i}\}$. The result $\nabla
_{u}v $ of the action of $\nabla _{u}$ on a tensor field $v\in \otimes
_{l}^{k}(M)$ is called covariant derivative of $v$ along $u$. For covariant
vectors and tensor fields an analogous relation holds, for instance, 
\begin{equation}
\nabla _{u}w=w_{i;j}\cdot u^{j}\cdot dx^{i}=(w_{i,j}+P_{ij}^{l}.w_{l})\cdot
u^{j}\cdot .dx^{i}\text{ \ , \ \ \ }w\in T^{\ast }(M)\text{ .}  \label{0.2}
\end{equation}
where $P_{ij}^{l}$ are the components of the covariant affine connection $P$
in a covariant co-ordinate basis $\{dx^{i}\}$. For $(L_{n},g)$, $Y_{n}$, $%
W_{n}$, $U_{n}$, and $V_{n}$-spaces $P_{ij}^{l}=-\Gamma _{ij}^{l}$.

$\pounds _{u}$ is the Lie differential operator \cite{Manoff-1} acting on
the elements of the tensor algebra $\mathcal{T}$ over $M$. The action of \ $%
\pounds _{u}$ is called dragging-along a contravariant vector field $u$. The
result $\pounds _{u}v$ of the action of $\pounds _{u}$ on a tensor field $v$
is called Lie derivative of $v$ along $u$.

The $n$-form $d\omega :=\frac{1}{n!}\cdot \sqrt{-d_{g}}\cdot \varepsilon
_{i_{1}...i_{n}}\cdot dx^{i_{1}}\wedge ...\wedge x^{i_{n}}$, where $%
d_{g}:=\det (g_{ij})<0$, $\varepsilon _{i_{1}...i_{n}}$ are the components
of the full antisymmetric Levi-Civita symbol, is called invariant volume
element in $M$.

The result of the action of the covariant differential operator $\nabla _{u}$
on the invariant volume element $d\omega $ can be written in the form \cite
{Manoff-1}, 
\begin{equation}
\nabla _{u}(d\omega )=\frac{1}{2}\cdot \overline{g}[\nabla _{u}g]\cdot
d\omega =\frac{1}{2}\cdot g^{\overline{i}\overline{j}}\cdot g_{ij;k}\cdot
u^{k}\cdot d\omega \text{ \ \ \ .}  \label{0.3}
\end{equation}

The result of the action of the Lie differential operator $\pounds _{u}$ on
the invariant volume element $d\omega $ can be written in the form \cite
{Manoff-1} 
\begin{equation}
\pounds _{u}(d\omega )=\frac{1}{2}\cdot \overline{g}[\pounds _{u}g]\cdot
d\omega =\frac{1}{2}\cdot g^{\overline{i}\overline{j}}\cdot \pounds
_{u}g_{ij}\cdot d\omega \text{ \ \ .}  \label{0.4}
\end{equation}

Let us recall some well known facts about differential-geometric structures
over spaces with affine connections and metrics.

\section{Recurrent relations in spaces with affine connections and metrics}

1. A parallel transport of a contravariant vector field $\xi \in T(M)$ along
the vector field $u$ could be defined in the form 
\begin{equation}
\nabla _{u}\xi =f\cdot \xi \text{ , \thinspace \thinspace \thinspace
\thinspace \thinspace \thinspace \thinspace \thinspace }f\in C^{r}(M)\text{ .%
}  \label{3.1}
\end{equation}

An equation of this type is called recurrent equation (or recurrent relation
for the vector field $\xi $) \cite{Sinjukov}, \cite{Manoff-7}

A special case of a parallel transport is the auto-parallel transport of a
contravariant vector field $u\in T(M)$ along itself 
\begin{equation}
\nabla _{u}u=k\cdot u\text{ , \thinspace \thinspace \thinspace \thinspace
\thinspace \thinspace \thinspace \thinspace }k\in C^{r}(M)\text{ .}
\label{1.1a}
\end{equation}

The recurrent relation (\ref{1.1a}) is the auto-parallel equation for $u$ in
its non-canonical form. After changing the parameter of the curve on which $%
u $ is a tangent vector the auto-parallel equation could be written in its
canonical form as \cite{Manoff-7} 
\begin{equation}
\nabla _{\overline{u}}\overline{u}=0\text{ \ \ .}  \label{1.1b}
\end{equation}

On the other side, after contracting (\ref{1.1a}) with $g(u)$, it follows
that 
\begin{eqnarray}
g(u,\nabla _{u}u) &=&k\cdot g(u,u):=k\cdot e\text{ \ \ , \ \ \ \ \ \ \ \ \ }%
g(u,u):=e:\neq 0\text{ \ \ ,}  \label{1.1c} \\
k &=&\frac{1}{e}\cdot g(u,\nabla _{u}u)=\frac{1}{2\cdot e}\cdot \lbrack
ue-(\nabla _{u}g)(u,u)]\text{ \ .}  \notag
\end{eqnarray}

For (pseudo) Riemannian spaces [$\nabla _{u}g=0$ for $\forall u\in T(M)$]
and normalized vector field $u$ ($e:=$const., $ue=0$) the function $k$ is
equal to zero and $\nabla _{u}u=0$.

2. The results of the action of the covariant differential operator $\nabla
_{u}$ and of the Lie differential operator $\pounds _{u}$ on the invariant
volume element $d\omega $ are recurrent relations for $d\omega $ 
\begin{equation}
\nabla _{u}(d\omega )=\overline{f}_{u}\cdot d\omega \text{ \ , \ \ \ \ }%
\overline{f}_{u}:=\frac{1}{2}\cdot \overline{g}[\nabla _{u}g]\in \otimes
_{0}^{0}(M)\subset C^{r}(M)\text{ ,}  \label{1.2}
\end{equation}
\begin{equation}
\pounds _{u}(d\omega )=\widetilde{f}_{u}\cdot d\omega \text{ \ , \ \ \ \ \ \
\ }\widetilde{f}_{u}:=\frac{1}{2}\cdot \overline{g}[\pounds _{u}g]\in
\otimes _{0}^{0}(M)\subset C^{r}(M)\text{ .}  \label{1.3}
\end{equation}

3. A conformal Killing vector field is defined by analogous type of a
recurrent equation.

\textit{Definition 1}. A conformal Killing vector field is a contravariant
vector field $u$ obeying the equation 
\begin{equation}
\pounds _{u}g=\lambda _{u}\cdot g\text{ \ \ \ , \ \ \ \ \ \ \ }\lambda
_{u}\in \otimes _{0}^{0}(M)\subset C^{r}(M)\text{ .}  \label{1.4}
\end{equation}

The equation is called conformal Killing equation. After contracting $%
\pounds _{u}g$ and $g$ from the last equation with $\overline{g}=g^{ij}\cdot
\partial _{i}.\partial _{j}$ [$\partial _{i}.\partial _{j}:=\frac{1}{2}\cdot
(\partial _{i}\otimes \partial _{j}+\partial _{j}\otimes \partial _{i})$] by
both basic vector fields, i.e. after finding the relations 
\begin{equation}
\overline{g}[\pounds _{u}g]=g^{\overline{k}\overline{l}}\cdot \pounds
_{u}g_{ij}\text{ \ \ \ \ \ \ and \ \ \ \ \ \ }\overline{g}[g]=g^{\overline{k}%
\overline{l}}.g_{kl}=n=\dim M\text{ \ ,}  \label{1.5}
\end{equation}
it follows for $\lambda _{u}$%
\begin{equation}
\lambda _{u}=\frac{1}{n}\cdot \overline{g}[\pounds _{u}g]=\frac{2}{n}\cdot 
\widetilde{f}_{u}\text{ .}  \label{1.6}
\end{equation}

Therefore, the factor $\lambda _{u}$ in a conformal Killing vector equation
is related to the factor $\widetilde{f}_{u}$ by which the invariant volume
element changes when dragged along the same vector field.

4. Let us now consider the condition for the existence of a Weyl's space.

\textit{Definition 2. }A Weyl's space is a differentiable manifold $M$ with $%
\dim M:=n$, provided with affine connections $\Gamma $ and $P$ (with $P\neq
-\Gamma $ or $P=-\Gamma $) and a metric $g$ with covariant derivative of $g$
along an arbitrary given contravariant vector field $u\in T(M)$ in the form 
\begin{equation}
\nabla _{u}g:=\frac{1}{n}\cdot Q_{u}\cdot g\text{ .}  \label{1.7}
\end{equation}
The existence condition is a recurrent relation for the metric $g$. Here 
\begin{eqnarray}
Q_{u} &=&Q_{j}\cdot u^{j}\text{ ,\thinspace\ \ \ \ }Q:=Q_{j}\cdot dx^{j}%
\text{ ,\ \ \ \ }  \label{1.8} \\
\text{\ }Q_{j} &:&=\underline{Q}_{k}\cdot f^{k}\,_{j}:=\underline{Q}_{%
\overline{j}}\text{ for }P\neq -\Gamma  \notag \\
Q_{j} &\equiv &Q_{j}\text{ for }P=-\Gamma \text{ .}  \notag
\end{eqnarray}

The covariant vector field (1-form) $\overline{Q}:=\frac{1}{n}\cdot
Q_{j}\cdot dx^{j}=\frac{1}{n}\cdot Q$ is called Weyl's covariant (covector)
field. If $Q$ is an exact form, i. e. if $Q=-d\overline{\varphi }=-\overline{%
\varphi }_{,j}\cdot dx^{j}$ with $Q_{j}=-\overline{\varphi }_{,j}$, $%
\overline{\varphi }\in C^{r}(M)$, $r\geqq 2$, then for a contravariant
vector field $u:=d/d\tau $ the invariant $Q_{u}$ could be written in the
form $Q_{u}=-d\overline{\varphi }/d\tau $. The scalar field $\overline{%
\varphi }$ is called dilaton field. The reason for this notation follows
from the properties of the Weyl's spaces considered below.

After contracting $\nabla _{u}g$ and $g$ from the last equation with $%
\overline{g}=g^{ij}\cdot \partial _{i}.\partial _{j}$ by both basic vector
fields, i.e. after finding the relations 
\begin{equation}
\overline{g}[\nabla _{u}g]=g^{\overline{i}\overline{j}}\cdot g_{ij;k}\cdot
u^{k}\text{ \ \ \ \ \ \ \ and \ \ \ \ \ \ \ }\overline{g}[g]=g^{\overline{k}%
\overline{l}}.g_{kl}=n=\dim M\text{ \ ,}  \label{1.9}
\end{equation}
it follows for $Q_{u}$%
\begin{equation}
Q_{u}=\overline{g}[\nabla _{u}g]=2\cdot \text{\ }\overline{f}_{u}\text{ \ ,
\ \ \ \ \ }\overline{f}_{u}=\frac{1}{2}\cdot Q_{u}\text{ \ \ \ \ \ .}
\label{1.10}
\end{equation}

Therefore, for Weyl's spaces, we have the recurrent relation for the
invariant volume element $d\omega $%
\begin{equation}
\nabla _{u}(d\omega )=\frac{1}{2}\cdot Q_{u}.d\omega \text{ \ \ \ \ .}
\label{1.11}
\end{equation}

5. If we now compare the recurrent relations obtained as a result of the
action of the covariant differential operator $\nabla _{u}$ and of the Lie
differential operator $\pounds _{u}$ on the invariant volume element $%
d\omega $ the question could arise what are the conditions for the
equivalence of the action of both the differential operators on $d\omega $,
i.e. under which conditions for the vector field $u$ and the metric $g$ we
could have the relation 
\begin{equation}
\nabla _{u}(d\omega )=\pounds _{u}(d\omega )\text{ ,}  \label{1.12}
\end{equation}
which is equivalent to the relation 
\begin{equation}
\overline{g}[\nabla _{u}g]=\overline{g}[\pounds _{u}g]\text{ \ \ \ \ or \ \
\ \ }\overline{g}[\nabla _{u}g-\pounds _{u}g]=0\text{ \ .}  \label{1.13}
\end{equation}

What does this relation physically mean? The action of the covariant
differential operator $\nabla _{u}$ is determined only on a curve at which $%
u $ is a tangent vector. The transport of $g$ is only on the curve and not
on the vicinities out of the points of the curve. The action of the Lie
differential operator is determined on the vicinities on and out of the
points of the curve with tangent vector $u$ on it. The dragging-along of $g$
is on the whole vicinities of the points of the curve and not only along the
curve. If the dragging-along $u$ of $g$ is equal to the transport of $g$
along $u$, then an observer with its worldline as the curve with tangent
vector $u$ could not observe any changes in its worldline vicinity different
from those who could register on its worldline. The observer will see its
surroundings as if they are moving with him.

It is obvious that in the general case, in $(\overline{L}_{n},g)$-spaces, a
sufficient condition for fulfilling the last relation is the condition 
\begin{equation}
\nabla _{u}g-\pounds _{u}g=0\text{ \ .}  \label{1.14}
\end{equation}

The last condition is fulfill:

(a) in Riemannian spaces (with or without torsion) [for which $\nabla _{u}g=0
$ for $\forall u\in T(M)$], when the Killing equation \cite{Yano} 
\begin{equation}
\pounds _{u}g=0\text{ \ }  \label{1.15}
\end{equation}
is fulfilled for the vector field $u$.

(b) in Weyl's spaces (with or without torsion) [for which $\nabla _{u}g=%
\frac{1}{n}\cdot Q_{u}\cdot g$ for $\forall u\in T(M)$], when the conformal
Killing equation 
\begin{equation}
\pounds _{u}g=\lambda _{u}\cdot g\text{ \ \ \ \ \ \ with \ \ \ \ \ \ }%
\lambda _{u}=\frac{1}{n}\cdot Q_{u}  \label{1.16}
\end{equation}
is fulfilled for the vector field $u$.

In $(\overline{L}_{n},g)$-spaces, the relation $\nabla _{u}g-\pounds _{u}g=0$
could be written in a co-ordinate basis in the form 
\begin{eqnarray}
\pounds _{u}g_{ij} &=&g_{ij;k}\cdot u^{k}+g_{kj}\cdot u^{\overline{k}}\,_{;%
\underline{i}}+g_{ik}\cdot u^{\overline{k}}\,_{;\underline{j}}+(g_{kj}\cdot
T_{l\underline{i}}\,^{\overline{k}}+g_{ik}\cdot T_{l\underline{j}}\,^{%
\overline{k}})\cdot u^{l}=  \notag \\
&=&g_{ij;k}\cdot u^{k}\text{ ,}  \label{1.17}
\end{eqnarray}
or in the forms 
\begin{equation}
g_{kj}\cdot u^{\overline{k}}\,_{;\underline{i}}+g_{ik}\cdot u^{\overline{k}%
}\,_{;\underline{j}}+(g_{kj}\cdot T_{l\underline{i}}\,^{\overline{k}%
}+g_{ik}\cdot T_{l\underline{j}}\,^{\overline{k}})\cdot u^{l}=0\text{ \ ,}
\label{1.18}
\end{equation}
\begin{equation}
g_{kj}\cdot (u^{\overline{k}}\,_{;\underline{i}}-T_{\underline{i}l}\,^{%
\overline{k}}\cdot u^{l})+g_{ik}\cdot (u^{\overline{k}}\,_{;\underline{j}%
}-T_{\underline{j}l}\,^{\overline{k}}\cdot u^{l})=0\text{ ,}  \label{1.19}
\end{equation}
where 
\begin{equation*}
T_{\underline{i}l}\,^{\overline{k}}:=f_{i}\,^{m}\cdot T_{ml}\,^{n}\cdot
f^{k}\,_{n}\text{, \ \ \ \ \ \ \ \ \ \ \ \ \ \ }\ \ u^{\overline{k}}\,_{;%
\underline{j}}:=f^{k}\,_{l}\cdot u^{l}\,_{;m}\cdot f_{j}\,^{m}.
\end{equation*}

The equation (\ref{1.19}) could be called ''generalized conformal Killing
equation'' for the vector field $u$ in the case $\pounds _{u}g=\nabla _{u}g$.

After multiplication of the last expression, equivalent to $\pounds
_{u}g_{ij}=g_{ij;k}\cdot u^{k}$, with $u^{\overline{i}}$ and $g^{m\overline{j%
}}$ and summation over $\overline{i}$ and $\overline{j}$ (and then changing
the index $m$ with $i$) we can find the equation for the vector field $u$ in
a co-ordinate basis 
\begin{equation}
u^{i}\,_{;j}\cdot u^{j}+g_{\overline{l}\overline{k}}\cdot u^{l}\cdot
(u^{k}\,_{;j}-T_{jm}\,^{k}\cdot u^{m})\cdot g^{ji}=0\text{ \ ,}  \label{1.20}
\end{equation}
or as an equation for the acceleration $a=a^{i}\cdot \partial _{i}$ with 
\begin{equation}
a^{i}=-g_{\overline{l}\overline{k}}\cdot u^{l}\cdot
(u^{k}\,_{;j}-T_{jm}\,^{k}\cdot u^{m})\cdot g^{ji}=-g_{\overline{l}\overline{%
k}}\cdot u^{l}\cdot k^{ki}\text{ \ ,}  \label{1.21}
\end{equation}
where 
\begin{equation}
k^{ki}=(u^{k}\,_{;j}-T_{jm}\,^{k}\cdot u^{m})\cdot g^{ji}\text{ .}
\label{1.22}
\end{equation}

On the other side, from the equation (\ref{1.19}) it is obvious that a
sufficient condition for fulfilling the equation (\ref{1.19}) is the
condition for $u^{i}$%
\begin{equation}
u^{k}\,_{;j}-T_{jl}\,^{k}\cdot u^{l}=0\text{ \ \ \ \ \ \ \ or \ \ \ \ \ }%
u^{k}\,_{;j}=T_{jl}\,^{k}\cdot u^{l}\text{ \ \ .\ }  \label{1.23}
\end{equation}

From the last expression, it follows that if the vector $u$ fulfills this
condition it should be an auto-parallel vector field since 
\begin{equation}
u^{i}\,_{;j}\cdot u^{j}=a^{i}=0\text{ \ \ \ .}  \label{1.24}
\end{equation}

If (\ref{1.23}) \ is fulfilled, then the following relations are valid: 
\begin{eqnarray*}
u^{m}\cdot R_{\,\,\;mkl}^{i}
&=&-(T_{ml}\,^{i}\,_{;k}-T_{mk}\,^{i}\,_{;l}+T_{ml}\,^{n}\cdot T_{nk}\,^{i}-
\\
&&-T_{mk}\,^{n}\cdot T_{nl}\,^{i}+T_{kl}\,^{n}\cdot T_{mn}\,^{i})\cdot u^{m}%
\text{ \ ,}
\end{eqnarray*}
\begin{equation*}
R_{mk}\cdot u^{m}\cdot u^{k}=g_{i}^{l}\cdot R^{i}\,_{mkl}\cdot u^{m}\cdot
u^{k}=T_{lm}\,^{l}\,_{;k}\cdot u^{k}\cdot u^{m}\text{ \ \ .}
\end{equation*}

Before going on to the kinematic characteristics of a vector field $u$
fulfilling the relations (\ref{1.19}) - (\ref{1.21}) or (\ref{1.23}) - (\ref
{1.24}), let us now consider some properties of a Weyl's space related to
the properties of scalar and vector fields in this type of spaces.

\section{Properties of a Weyl's space}

1. Parallel transports over Weyl's spaces are at the same time conformal
transports. This means that if $\nabla _{u}\xi =0$ and $\nabla _{u}\eta =0$,
then $ul_{\xi }=(1/2n)\cdot Q_{u}\cdot l_{\xi }$, $ul_{\eta }=(1/2n)\cdot
Q_{u}\cdot l_{\eta }$, and \ $u[\cos (\xi ,\eta )]=0$, where $l_{\xi }:=\mid
g(\xi ,\xi )\mid ^{1/2}$, $l_{\eta }:=\;\mid g(\eta ,\eta )\mid ^{1/2}$, $%
\cos (\xi ,\eta ):=[g(\xi ,\eta )]/(l_{\xi }\cdot l_{\eta })$. If $u=d/d\tau 
$, then \cite{Manoff-e} 
\begin{equation}
\frac{dl_{\xi }}{d\tau }=\frac{1}{2n}\cdot Q_{u}\cdot l_{\xi }\text{ , \ \ \
\ \ \ }\frac{dl_{\eta }}{d\tau }=\frac{1}{2n}\cdot Q_{u}\cdot l_{\eta }\text{%
\ \ \ \ ,}  \label{1.25}
\end{equation}
and therefore, 
\begin{eqnarray}
l_{\xi } &=&l_{\xi 0}\cdot \exp [\frac{1}{2n}\cdot \int Q_{j}\cdot dx^{j}]%
\text{ , \ }l_{\eta }=l_{\eta 0}\cdot \exp [\frac{1}{2n}\cdot \int
Q_{j}\cdot dx^{j}]\text{ , \ \ }  \label{1.26} \\
\text{\ }l_{\xi 0} &=&\text{const., }l_{\eta 0}=\text{const., \ \ \ \ \ }%
\cos (\xi ,\eta )=\text{const. along }\tau (x^{k})\text{. \ \ \ \ \ \ } 
\notag
\end{eqnarray}

Furthermore, if $Q_{j}=-n\cdot \overline{\varphi }_{,j}$ and respectively $%
Q_{u}=-n\cdot d\overline{\varphi }/d\tau $ , then the equation for $l_{\xi }$
obtains in this case the simple form 
\begin{equation}
\frac{dl_{\xi }}{d\tau }=-\frac{1}{2}\cdot \frac{d\overline{\varphi }}{d\tau 
}\cdot l_{\xi }\text{ .}  \label{1.27}
\end{equation}

The solution for $l_{\xi }$ could easily be found as 
\begin{equation}
l_{\xi }=l_{\xi 0}\cdot e^{-\frac{1}{2}\cdot \overline{\varphi }}\text{ \ \
\ .}  \label{1.27a}
\end{equation}

The scalar field $\overline{\varphi }$ [as an invariant function $\overline{%
\varphi }\in \otimes ^{0}\,_{0}(M)$] appears as a gauge factor changing the
length of the vector $\xi $. This is the reason for calling the scalar field 
$\overline{\varphi }$ \textit{dilaton field} in a Weyl's space.

2. \ The metric in a Weyl's space has properties which can be formulated in
the following two propositions:

\textit{Proposition 1}. \cite{Bonneau} A metric $\widetilde{g}$ conformal to
a Weyl's metric $g$ is also a Weyl's metric. In other words, if $\widetilde{g%
}=e^{\varphi }\cdot g$ with $\nabla _{\xi }g=\frac{1}{n}\cdot Q_{\xi }\cdot g
$ for $\forall \xi \in T(M)$, then $\ \nabla _{\xi }\widetilde{g}=\frac{1}{n}%
\cdot \widetilde{Q}_{\xi }\cdot \widetilde{g}$.

The proof is trivial.

Therefore, all Weyl's metrics belong to the set of all metrics conformal to
a Weyl's metric.

Let the square $ds^{2}$ of the line element $ds$ for a Weyl's metric $g$ in $%
\overline{W}_{n}$- (or $\overline{Y}_{n}$)-spaces and in $W_{n}$- (or $Y_{n}$%
)-spaces be given in the forms respectively 
\begin{equation*}
ds^{2}=g_{\overline{i}\overline{j}}\cdot dx^{i}\cdot dx^{j}\text{ \ \ , \ \
\ \ \ \ \ \ \ \ }ds^{2}=g_{ij}\cdot dx^{i}\cdot dx^{j}\text{ \ \ .}
\end{equation*}

Then, the square $d\widetilde{s}^{2}$ of the line element $d\widetilde{s}$
for a conformal to the Weyl's metric $g$ will have the forms respectivelly 
\begin{equation*}
d\widetilde{s}^{2}=\widetilde{g}_{\overline{i}\overline{j}}\cdot dx^{i}\cdot
dx^{j}=e^{\varphi }.ds^{2}\text{ \ \ , \ \ \ \ \ \ \ \ \ \ }d\widetilde{s}%
^{2}=\widetilde{g}_{ij}\cdot dx^{i}\cdot dx^{j}=e^{\varphi }.ds^{2}\text{ \
\ .}
\end{equation*}

The invariant function $\varphi =\varphi (x^{k})\in \otimes
_{0}^{0}(M)\subset C^{r}(M)$, $r\geqq 2$, is also called \textit{dilaton
field} because of its appearing as a gauge factor changing the line element
in a Weyl's space.

For $\pm \widetilde{l}_{d}^{2}=\widetilde{g}(d,d)=d\widetilde{s}^{2}=%
\widetilde{g}_{\overline{i}\overline{j}}\cdot dx^{i}\cdot dx^{j}$ with $%
d^{i}:=dx^{i}$ we have 
\begin{equation*}
\pm \widetilde{l}_{d}^{2}=d\widetilde{s}^{2}=e^{\varphi }.ds^{2}=\pm
e^{\varphi }.l_{d}^{2}\text{\ .}
\end{equation*}

On the other side, by the use of the expression (\ref{1.26}) for the length $%
\widetilde{l}_{d}$ we find that 
\begin{eqnarray*}
\widetilde{l}_{d}^{2} &=&\widetilde{l}_{d0}^{2}.\exp [\frac{1}{n}\cdot \int 
\widetilde{Q}_{j}\cdot dx^{j}]=\widetilde{l}_{d0}^{2}\cdot \exp [\frac{1}{n}%
\cdot \int (n\cdot \varphi _{,j}+Q_{j})\cdot dx^{j}= \\
&=&\widetilde{l}_{d0}^{2}\cdot \exp \varphi \cdot \exp [\frac{1}{n}\cdot
\int Q_{j}\cdot dx^{j}]=e^{\varphi }\cdot l_{d}^{2}\text{ \ , \ \ \ \ \ \ \
\ \ \ }\widetilde{l}_{d0}^{2}:=l_{d0}^{2}=\text{ const.}
\end{eqnarray*}

For $Q_{j}=-n\cdot \overline{\varphi }_{,j}$, it follows that 
\begin{equation*}
\widetilde{l}_{d}^{2}=\widetilde{l}_{d0}\cdot e^{\varphi -\overline{\varphi }%
}\text{ \ .\ \ }
\end{equation*}

If $\widetilde{l}_{d}^{2}$ does not change along the vector field $u$, then $%
\varphi =\overline{\varphi }$ and we have only one dilation field $\varphi
=\varphi (x^{k})=\overline{\varphi }(x^{k})$ which determines the conformal
factor of the metric $\widetilde{g}$, conformal to a Weyl's metric $g$, as
well as the Weyl's covector $Q$. In this case the metric $\widetilde{g}$
appears as a Riemannian metric because of $\widetilde{Q}_{j}=0$. If $\varphi
\neq \overline{\varphi }$, then the metric $\widetilde{g}$, conformal to $g$
is again a Weyl's metric with $\widetilde{Q}_{j}=-n\cdot \widetilde{\varphi }%
_{,j}$ and with $\widetilde{\varphi }=-(\overline{\varphi }-\varphi )$.

\textit{Proposition} 2. The necessary and sufficient condition for a metric $%
\widetilde{g}$ conformal ($\widetilde{g}=e^{\varphi }\cdot g$) to a Weyl's
metric $g$ [obeying the condition $\nabla _{\xi }g=\frac{1}{n}Q_{\xi }\cdot g
$ for $\forall \xi \in T(M)$] to be a Riemannian metric $\widetilde{g}$
[obeying the condition $\nabla _{\xi }\widetilde{g}=0$ for $\forall \xi \in
T(M)$] is the condition 
\begin{equation}
Q_{\xi }=-n\cdot (\xi \varphi )\text{ \ \ \ \ \ , \ \ \ \ }\xi \in T(M)\text{
, \ \ \ \ \ }\varphi \in C^{r}(M)\text{ , }r\geqq 2\text{ .}  \label{1.29}
\end{equation}

The proof is trivial.

\textit{Corollary}. All Riemannian metrics are conformal to a Weyl's metric
in a Weyl's space with $Q_{\xi }=-n\cdot \xi \varphi $ for $\forall \xi \in
T(M)$, $\varphi \in C^{r}(M)$, $r\geqq 2$, [or in a co-ordinate basis with $%
Q_{k}=-n\cdot $\ $\varphi _{,k}$].

On the basis of the last proposition we can state that for \textit{every }%
given Riemannian metric $\widetilde{g}$ from a Riemannian space and a\textit{%
\ given} \textit{scalar (dilaton) field} $\varphi (x^{k})$ in this space we
could generate a Weyl's metric $g$ in a Weyl's space with the same affine
connection as the affine connection in the Riemannian space. Vice versa, for
every given Weyl's space with Weyl's covariant vector field $Q$ constructed
by a scalar (dilaton) field $\varphi $ and a Weyl's metric $g$ we could
generate a Riemannian metric $\widetilde{g}$ in a Riemannian space with the
same affine connection as in the corresponding Weyl's space.

Therefore, \textit{every (metric) tensor-scalar theory of gravitation in a
(pseudo) Riemannian space (with or without torsion) could be reformulated in
a corresponding Weyl's space with Weyl's metric and dilaton field,
generating the Weyl's covector in the Weyl's space} and vice versa: \textit{%
every (metric) tensor-scalar theory in a Weyl's space with scalar (dilaton)
field, generating the Weyl's covector, could be reformulated in terms of a
(metric) tensor-scalar theory in the corresponding Riemannian space with the
same affine connections as the affine connection in the Weyl's space.}

The kinematic characteristics of a vector field obeying the condition $%
\nabla _{u}g-\pounds _{u}g=0$ can now be considered from a more general
point of view, namely, for spaces with affine connections (which components
differ not only by sign) and then specialized for Weyl's spaces.

\subsection{Deformation velocity, shear velocity, rotation velocity and
expansion velocity}

The relative velocity 
\begin{eqnarray}
_{rel}v &=&\overline{g}[h_{u}(\nabla _{u}\xi )]=g^{ij}\cdot h_{\overline{j}%
\overline{k}}\cdot \xi ^{k}\,_{;l}\cdot u^{l}\cdot e_{i}\text{ \ \ \ ,}
\label{2.11} \\
\,\,\,\,e_{i} &=&\partial _{i}\text{ (in a co-ordinate basis),}  \notag
\end{eqnarray}
where $\overline{g}[h_{u}(\xi )]:=\xi _{\perp }=g^{ik}\cdot h_{\overline{k}%
\overline{l}}\cdot \xi ^{l}\cdot e_{i}\,$ is called deviation vector field
and (the indices in a co-ordinate and in a non-co-ordinate basis are written
in both cases as Latin indices instead of Latin and Greek indices)

\begin{equation}
\begin{array}{c}
h_{u}=g-\frac{1}{e}.g(u)\otimes g(u)\text{ , \ \ \ \ \ \ \ }%
h_{u}=h_{ij}\cdot e^{i}.e^{j}\text{ , \ \ \ \ \ \ \ \ \ \ \ \ \ \ \ }%
\overline{g}=g^{ij}\cdot e_{i}.e_{j}, \\ 
\nabla _{u}\xi =\xi ^{i}\text{ }_{;j}\cdot u^{j}\cdot e_{i}\text{ ,
\thinspace \thinspace \thinspace \thinspace \thinspace \thinspace \thinspace
\thinspace \thinspace \thinspace \thinspace \thinspace \thinspace \thinspace
\thinspace \thinspace }\xi ^{i}\text{ }_{;j}=e_{j}\xi ^{i}+\Gamma
_{kj}^{i}\cdot \xi ^{k}\,\text{,\thinspace \thinspace \thinspace \thinspace
\thinspace \thinspace \thinspace \thinspace \thinspace \thinspace \thinspace
\thinspace \thinspace \thinspace \thinspace \thinspace \thinspace \thinspace
\thinspace }\,\,\,\Gamma _{kj}^{i}\neq \Gamma _{jk}^{i}\text{ }, \\ 
e=g(u,u)=g_{\overline{i}\overline{j}}\cdot u^{i}\cdot u^{j}=u_{\overline{i}%
}\cdot u^{i}\neq 0\text{ ,\thinspace \thinspace \thinspace \thinspace
\thinspace \thinspace \thinspace \thinspace \thinspace \thinspace \thinspace
\thinspace \thinspace \thinspace }g(u)=g_{\overline{i}\overline{k}}\cdot
u^{k}\cdot e^{i}=u_{\overline{i}}\cdot e^{i}\text{ ,} \\ 
\text{\thinspace }h_{ij}=g_{ij}-\frac{1}{e}\cdot u_{i}\cdot u_{j}\text{ ,}
\\ 
h_{u}(\nabla _{u}\xi )=h_{i\overline{j}}\cdot \xi ^{j}\text{ }_{;k}\cdot
u^{k}\cdot e^{i}\text{ .}
\end{array}
\label{2.4}
\end{equation}
could be written in a $(\overline{L}_{n},g)$-space under the conditions $%
g(u,\xi ):=l=0$, $\pounds _{\xi }u=0$, in the form \cite{Manoff-9}, \cite
{Manoff-9a}
\begin{equation*}
_{rel}v=\overline{g}[d(\xi )]\text{ .}
\end{equation*}

The covariant tensor field $d$ is a generalization for $(\overline{L}_{n},g)$%
-spaces of the well known 
\index{deformation@deformation!deformation velocity@deformation velocity} 
\textit{deformation velocity }tensor for $V_{n}$-spaces \cite{Stephani}, 
\cite{Kramer}. It is usually represented by means of its three parts: the
trace-free symmetric part, called \textit{shear velocity }tensor (shear),
the anti-symmetric part, called 
\index{rotation@rotation!rotation velocity@rotation velocity} \textit{%
rotation velocity }tensor (rotation) and the trace part, in which the trace
is called 
\index{expansion@expansion!expansion velocity@expansion velocity} \textit{%
expansion velocity }(expansion)\textit{\ }invariant.

After some more complicated as for $V_{n}$-spaces calculations, the
deformation velocity tensor $d$ can be given in the form \cite{Manoff-9}

\begin{equation}
d=h_{u}(k)h_{u}=h_{u}(k_{s})h_{u}+h_{u}(k_{a})h_{u}=\sigma +\omega +%
\frac{1}{n-1}\cdot \theta \cdot h_{u}\text{ ,}  \label{2.12}
\end{equation}

\noindent where
\begin{equation*}
\begin{array}{c}
k[g(\xi )]=\nabla _{\xi }u-T(\xi ,u)\text{ , \thinspace \thinspace
\thinspace }k=(u^{i}\text{ }_{;l}-T_{lk}\,^{i}.u^{k}).g^{lj}.e_{i}\otimes
e_{j}=k^{ij}.e_{i}\otimes e_{j}\text{ ,} \\ 
k[g(u)]=k(g)u=k^{ij}.g_{\overline{j}\overline{k}}.u^{k}.e_{i}=a=\nabla
_{u}u=u^{i}\text{ }_{;j}.u^{j}.e_{i}\text{ ,}
\end{array}
\end{equation*}
\begin{eqnarray*}
k_{s} &=&\,_{s}k^{ij}\cdot e_{i}.e_{j}\text{, }_{s}k^{ij}=\frac{1}{2}%
(k^{ij}+k^{ji})\text{,} \\
_{a}k &=&\,_{a}k^{ij}\cdot e_{i}\wedge e_{j},_{a}k^{ij}=\frac{1}{2}%
(k^{ij}-k^{ji}),e_{i}\wedge e_{j}=\frac{1}{2}(e_{i}\otimes
e_{j}-e_{j}\otimes e_{i})\text{ .}
\end{eqnarray*}

The tensor $\sigma $ is the 
\index{shear@shear!shear velocity@shear velocity} \textit{shear velocity}
tensor (shear) , 
\begin{equation}
\begin{array}{c}
\sigma =\,_{s}E-\,_{s}P=E-P-%
\frac{1}{n-1}\cdot \overline{g}[E-P]\cdot h_{u}=\sigma _{ij}\cdot
e^{i}.e^{j}= \\ 
=E-P-\frac{1}{n-1}\cdot (\theta _{o}-\theta _{1})\cdot h_{u}\text{ ,}
\end{array}
\label{2.13}
\end{equation}

\noindent where

\begin{equation}
\begin{array}{c}
_{s}E=E-\frac{1}{n-1}\cdot \overline{g}[E]\cdot h_{u}\text{ , \thinspace
\thinspace \thinspace \thinspace \thinspace \thinspace \thinspace \thinspace
\thinspace \thinspace \thinspace \thinspace \thinspace \thinspace }\overline{%
g}[E]=g^{\overline{i}\overline{j}}\cdot E_{ij}=\theta _{o}\text{ ,} \\ 
E=h_{u}(\varepsilon )h_{u}\text{ , \thinspace \thinspace \thinspace
\thinspace \thinspace \thinspace \thinspace \thinspace \thinspace \thinspace
\thinspace \thinspace }k_{s}=\varepsilon -m\text{ , \thinspace \thinspace
\thinspace \thinspace \thinspace \thinspace \thinspace \thinspace \thinspace
\thinspace \thinspace }\varepsilon =\frac{1}{2}(u_{\text{ };l}^{i}\cdot
g^{lj}+u_{\text{ };l}^{j}\cdot g^{li})\cdot e_{i}.e_{j}\text{ ,} \\ 
m=\frac{1}{2}(T_{lk}\,^{i}\cdot u^{k}\cdot g^{lj}+T_{lk}\,^{j}\cdot
u^{k}\cdot g^{li})\cdot e_{i}.e_{j}\text{ .}
\end{array}
\label{2.14}
\end{equation}

The tensor $_{s}E$ is the \textit{torsion-free shear velocity} 
\index{shear@shear!torsion-free shear velocity@torsion-free shear velocity}%
\textit{\ }tensor, the tensor $_{s}P$ is the 
\index{shear@shear!shear velocity induced by the torsion@shear velocity induced by the torsion}
\textit{shear velocity} tensor \textit{induced by the torsion},

\textit{
\begin{equation}
\begin{array}{c}
_{s}P=P-%
\frac{1}{n-1}\cdot \overline{g}[P]\cdot h_{u}\text{ , \thinspace \thinspace
\thinspace \thinspace \thinspace \thinspace \thinspace \thinspace \thinspace
\thinspace \thinspace \thinspace \thinspace \thinspace \thinspace }\overline{%
g}[P]=g^{\overline{k}\overline{l}}\cdot P_{kl}=\theta _{1}\text{, \thinspace
\thinspace \thinspace \thinspace \thinspace \thinspace }P=h_{u}(m)h_{u}\text{
,} \\ 
\theta _{1}=T_{kl}\,^{k}\cdot u^{l}\text{ ,\thinspace \thinspace \thinspace
\thinspace \thinspace \thinspace \thinspace }\theta _{o}=u^{n}\text{ }_{;n}-%
\frac{1}{2e}(e_{,k}\cdot u^{k}-g_{kl;m}\cdot u^{m}\cdot u^{\overline{k}%
}\cdot u^{\overline{l}})\text{ ,\thinspace \thinspace \thinspace }\theta
=\theta _{o}-\theta _{1}\text{ . }
\end{array}
\label{2.15}
\end{equation}
}

The invariant $\theta $ is the 
\index{expansion@expansion!expansion velocity@expansion velocity} \textit{%
expansion velocity,} the invariant\textit{\ $\theta _{o}$} is the 
\index{expansion@expansion!torsion-free expansion velocity@torsion-free expansion velocity}
\textit{torsion-free expansion velocity,} the invariant $\theta _{1}$ is the 
\textit{expansion velocity induced by the torsion.}

The tensor $\omega $ is the 
\index{rotation@rotation!rotation velocity@rotation velocity}\textit{\
rotation velocity }tensor (rotation velocity),

\begin{equation}
\begin{array}{c}
\omega =h_{u}(k_{a})h_{u}=h_{u}(s)h_{u}-h_{u}(q)h_{u}=S-Q%
\text{ ,} \\ 
s=\frac{1}{2}(u^{k}\text{ }_{;m}\cdot g^{ml}-u^{l}\text{ }_{;m}\cdot
g^{mk})\cdot e_{k}\wedge e_{l}\text{ ,} \\ 
q=\frac{1}{2}(T_{mn}\,^{k}\cdot g^{ml}-T_{mn}\,^{l}\cdot g^{mk})\cdot
u^{n}\cdot e_{k}\wedge e_{l}\text{ , \ \ } \\ 
\text{\thinspace \thinspace }S=h_{u}(s)h_{u}\text{ , \thinspace \thinspace }%
Q=h_{u}(q)h_{u}\text{ .}
\end{array}
\label{2.16}
\end{equation}

The tensor $S$ is the 
\index{rotation@rotation!torsion-free rotation velocity@torsion-free rotation velocity}
\textit{torsion-free rotation velocity} tensor, the tensor $Q$ is the 
\index{rotation@rotation!rotation velocity induced by the torsion@rotation velocity induced by the torsion}%
\textit{\ rotation velocity }tensor \textit{induced by the torsion.}

By means of the expressions for $\sigma $, $\omega $ and $\theta $ the
deformation velocity tensor $d$ can be decomposed in two parts: $d_0$ and $%
d_1$

\begin{equation}
d=d_{o}-d_{1}%
\text{ , \ \ \ \thinspace \thinspace \thinspace }d_{o}=\,_{s}E+S+\frac{1}{n-1%
}.\theta _{o}.h_{u}\text{ , \ \ \thinspace \thinspace \thinspace \thinspace }%
d_{1}=\,_{s}P+Q+\frac{1}{n-1}.\theta _{1}.h_{u}\text{ ,}  \label{2.17}
\end{equation}

\noindent where $d_o$ is the 
\index{deformation@deformation!torsion-free deformation velocity@torsion-free deformation velocity}
\textit{torsion-free deformation velocity} tensor and $d_1$ is the 
\index{deformation@deformation!deformation velocity induced by the torsion@deformation velocity induced by the torsion}
\textit{deformation velocity }tensor \textit{induced by the torsion. }For
the case of $V_n$-spaces $d_1=0$ ($_sP=0$, $Q=0$, $\theta _1=0$).

After some calculations, the shear velocity tensor $\sigma $ and the
expansion velocity $\theta $ can also be written in the forms

\begin{equation}
\begin{array}{c}
\sigma =%
\frac{1}{2}\{h_{u}(\nabla _{u}\overline{g}-\pounds _{u}\overline{g})h_{u}-%
\frac{1}{n-1}\cdot (h_{u}[\nabla _{u}\overline{g}-\pounds _{u}\overline{g}%
])\cdot h_{u}\}\text{ }= \\ 
\\ 
=\frac{1}{2}\{h_{i\overline{k}}\cdot (g^{kl}\text{ }_{;m}\cdot u^{m}-\pounds
_{u}g^{kl})\cdot h_{\overline{l}j}-\frac{1}{n-1}\cdot h_{\overline{k}%
\overline{l}}\cdot (g^{kl}\text{ }_{;m}\cdot u^{m}-\pounds _{u}g^{kl})\cdot
h_{ij}\}\cdot e^{i}.e^{j}\text{ ,}
\end{array}
\label{2.18}
\end{equation}
\begin{equation}
\theta =\frac{1}{2}\cdot h_{u}[\nabla _{u}\overline{g}-\pounds _{u}\overline{%
g}]=\frac{1}{2}h_{\overline{i}\overline{j}}\cdot (g^{ij}\text{ }_{;k}\cdot
u^{k}-\pounds _{u}g^{ij})\text{ .}  \label{2.19}
\end{equation}

The physical interpretation of the velocity tensors $d$, $\sigma $, $\omega $%
, and of the invariant $\theta $ for the case of $V_{4}$-spaces \cite{Synge}%
, 
\index{Synge J. L.@Synge J. L.} \cite{Ehlers} 
\index{Ehlers J.@Ehlers J.}, can also be extended for $(%
\overline{L}_{4},g)$-spaces. It is easy to be seen that the existence of
some kinematic characteristics ($_{s}P$, $Q$, $\theta _{1}$) depends on the
existence of the torsion tensor field. They vanish if it is equal to zero
(e.g. in $V_{n}$-spaces). It should be stressed that the decomposition of
the deformation tensor $d$ could not follow from the decomposition of $%
u^{i}\,_{;j}$ as this has been done by Ehlers \cite{Ehlers} for (pseudo)
Riemannian spaces without torsion ($V_{n}$-spaces, $n=4$). In $(\overline{L}%
_{n},g)$-spaces 
\begin{eqnarray}
u^{i}\,_{;j} &=&\frac{1}{e}\cdot a^{i}\cdot u_{\overline{j}}+g^{ik}\cdot
(_{s}E_{\overline{k}\overline{j}}+S_{\overline{k}\overline{j}}+\frac{1}{n-1}%
\cdot \theta _{0}\cdot h_{\overline{k}\overline{j}})+  \label{2.19a} \\
&&+\frac{1}{2\cdot e}\cdot u^{i}\cdot (e_{,l}-g_{mn;l}\cdot u^{\overline{m}%
}\cdot u^{\overline{n}})\cdot h^{lk}\cdot g_{\overline{k}\overline{j}}\text{
\ .}  \notag
\end{eqnarray}

The representation of the shear velocity tensor $\sigma $ and the expansion
invariant $\theta $ by means of the covariant and Lie derivatives of the
contravariant metric tensor $\overline{g}$ give rise to some important
conclusions about their vanishing or nonvanishing in a $(\overline{L}_{n},g)$%
-space.

From the structure of $\sigma $ and $\theta $ in the last two expression, it
is obviously that if $\pounds _{u}\overline{g}=\nabla _{u}\overline{g}$ then 
$\sigma =0$ and $\theta =0$, i.e. the condition $\pounds _{u}\overline{g}%
=\nabla _{u}\overline{g}$ appears as a sufficient conditions for $\sigma =0$
and $\theta =0$. On the other side, this condition could be written in the
form $\pounds _{u}g=\nabla _{u}g$ because of the relations $\pounds _{u}%
\overline{g}=-\overline{g}(\pounds _{u}g)\overline{g}$ and $\nabla _{u}%
\overline{g}=-\overline{g}(\nabla _{u}g)\overline{g}$. For Weyl's spaces ($%
\nabla _{u}g=\frac{1}{n}\cdot Q_{u}\cdot g$) the same condition degenerate
in the condition for the existence of a conformal Killing vector $u$%
\begin{equation}
\pounds _{u}g=\lambda _{u}\cdot g\text{ \ \ \ \ \ \ with \ \ \ \ }\lambda
_{u}=\frac{1}{n}\cdot Q_{u}\text{ .}  \label{2.20}
\end{equation}

On the basis of the above considerations we could now formulate the
following propositions:

\textit{Proposition 3}. If a metric $g$ in a space with affine connections
and metrics [a $(\overline{L}_{n},g)$- or a $(L_{n},g)$-space] fulfills the
condition 
\begin{equation}
\pounds _{u}\overline{g}=\nabla _{u}\overline{g}\text{ \ \ or \ \ \ \ \ \ }%
\pounds _{u}g=\nabla _{u}g\text{ , }  \label{2.21}
\end{equation}
then the space admits a non-null shear-free and expansion-free contravariant
vector field $u$.

\textit{Proposition 4}. If a contravariant non-null vector field fulfills in
a Weyl's space a conformal Killing equation of the type 
\begin{equation}
\pounds _{u}g=\lambda _{u}\cdot g\text{ \ \ \ \ \ \ with \ \ \ \ }\lambda
_{u}=\frac{1}{n}\cdot Q_{u}\text{ ,}  \label{2.22}
\end{equation}
then this conformal vector field $u$ is also a shear-free and expansion-free
vector field.

\textit{Proposition 5}. If a contravariant non-null vector field $u$ in a
space with affine connections and metrics [a $(\overline{L}_{n},g)$- or a $%
(L_{n},g)$-space] fulfills the equation (\ref{1.23}) 
\begin{equation*}
u^{k}\,_{;j}-T_{jl}\,^{k}\cdot u^{l}=0\text{ \ \ \ \ \ \ \ or \ \ \ \ \ }%
u^{k}\,_{;j}=T_{jl}\,^{k}\cdot u^{l}\text{ ,}
\end{equation*}
then this vector field is an auto-parallel shear-free and expansion-free
vector field.

\textit{Proposition 6}. If a contravariant non-null vector field $u$ in a
Weyl's space fulfills the equation (\ref{1.23}) 
\begin{equation*}
u^{k}\,_{;j}-T_{jl}\,^{k}\cdot u^{l}=0\text{ \ \ ,}
\end{equation*}
then it is an auto-parallel, shear-free and expansion-free conformal Killing
vector field.

The auto-parallel equation (\ref{1.1a}) for the vector field $u$ is
interpreted as an equation of motion for a free spinless test particle in
spaces with affine connection and metrics \cite{Manoff-6}, \cite{Manoff-7}.
Let us now consider this equation more closely in Weyl's spaces.

\section{Auto-parallel equation in Weyl's spaces as an equation for a free
moving spinless test particles}

Usually the following definition of a free moving test particle in a space
with affine connections and metrics [and especially in (pseudo) Riemannian
spaces without torsion] is introduced \cite{Borisova-1}:

\textit{Definition 3}. A free spinless test particle is a material point
with rest mass (density) $\rho $, velocity $u$ (as tangent vector $u$ to its
trajectory) and momentum (density) $p:=\rho \cdot u$ with the following
properties:

(a) The momentum density $p$ does not change its direction along the world
line of the material point, i.e. the vector $p$ fulfills the recurrent
condition $\nabla _{u}p=f\cdot p$, or the condition $\nabla _{u}p=0$, as
conditions for parallel transport along $u$.

(b) The momentum density $p$ does not change its length $l_{p}=\mid
g(p,p)\mid ^{1/2}$ along the world line of the material point, i.e. $p$
fulfills the condition $ul_{p}=0$.

The change of the length of a vector field $p$ along a vector field $u$ in a 
$(\overline{L}_{n},g)$-space could be found in the form \cite{Manoff-e} 
\begin{equation}
ul_{p}=\pm \frac{1}{2\cdot l_{p}}\cdot \lbrack (\nabla _{u}g)(p,p)+2\cdot
g(\nabla _{u}p,p)]\text{ \ \ , \ \ \ \ \ \ \ \ \ }l_{p}\neq 0\text{ \ .}
\label{4.1}
\end{equation}

Let us now consider the two conditions (a) and (b) for $p$ separately to
each other.

(a) If we write $p$ in its explicit form $p=\rho \cdot u$, then the
condition for a parallel transport of $p$ along $u$ could be written as 
\begin{equation}
\nabla _{u}u=[f-u(\log \rho )]\cdot u\text{ ,}  \label{4.2}
\end{equation}
with 
\begin{equation}
f=u(\log \rho )+\frac{1}{2\cdot e}\cdot \lbrack ue-(\nabla _{u}g)(u,u)]\text{
\ , \ \ \ \ \ \ }e=g(u,u)\neq 0\text{ ,}  \label{4.3}
\end{equation}
and 
\begin{equation}
\nabla _{u}u=\frac{1}{2\cdot e}\cdot \lbrack ue-(\nabla _{u}g)(u,u)]\cdot u%
\text{ .}  \tag{(a)}  \label{4.4}
\end{equation}

In the special case of \ (pseudo) Riemannian spaces ($V_{n}$- or $U_{n}$-
spaces), where $\nabla _{u}g=0$, $e=$ const. $\neq 0$, $\rho =\,$const., $%
f=0 $, it follows that $\nabla _{u}p=0$, and $\nabla _{u}u=0$. At the same
time, $ul_{p}=0$. The parallel equation $\nabla _{u}p=0$ has as a corollary
the preservation of the length $l_{p}$ of $p$ along $u$. This is not the
case if a space is not a (pseudo) Riemannian space.

(b) The conservation of the momentum density $p$ along the trajectory of the
particles $[ul_{p}=0]$ requires the transport of $p$ on this trajectory to
be of the type of a Fermi-Walker transport, i.e. $p$ should obey an equation
of the type \cite{Manoff-2}, \cite{Manoff-3} 
\begin{equation}
\nabla _{u}p=\overline{g}(^{F}\omega -\frac{1}{2}\cdot \nabla _{u}g)(p)=%
\overline{g}[^{F}\omega (p)]-\frac{1}{2}\cdot \overline{g}(\nabla _{u}g)(p)%
\text{ ,}  \label{3.3}
\end{equation}
where $^{F}\omega \in \Lambda ^{2}(M)$ is an antisymmetric tensor of 2nd
rank. For a free particle it could be related to the rotation velocity
tensor (\ref{2.16}) of the velocity $u$, i.e. $^{F}\omega :=\omega $. Then $%
\omega (p)=0$ [because of $\omega (\rho \cdot u)=\rho \cdot (\omega (u))=0$]
and we have for $\nabla _{u}p$%
\begin{equation}
\nabla _{u}p=-\frac{1}{2}\cdot \overline{g}(\nabla _{u}g)(p)\text{ , \ \ \ \
\ \ \ \ }ul_{p}=0\text{ \ .}  \label{3.4}
\end{equation}

\bigskip For the vector field $u$ follows the corresponding condition

\begin{equation}
\nabla _{u}u=-\{[u(\log \rho )]\cdot u+\frac{1}{2}\cdot \overline{g}(\nabla
_{u}g)(u)\}\text{ \ .}  \tag{(b)}  \label{3.5}
\end{equation}

Therefore, \ for $u$ we have two equations as corollaries from the
requirements for the momentum density $p$: equation (a) which follows from
the condition for preservation of the direction of the momentum density $p$,
and equation (b) which follows from the condition for preservation of the
length $l_{p}$ of the momentum density $p$.

(a) The first equation (a) \ for $u$ \ is the auto-parallel equation in its
non-canonical form. It does not depend on the rest mass density $\rho $ of
the particle. From the equation, it follows that the necessary and
sufficient condition for a spinless test particle to move in space with
affine connections and metrics on a trajectory described by the
auto-parallel equation in its canonical form $(\nabla _{u}u=0)$ is the
condition 
\begin{equation}
\lbrack ue-(\nabla _{u}g)(u,u)]\cdot u=0\text{ \ \ \ , \ \ \ \ \ \ }e\neq 0%
\text{ .}  \label{4.7}
\end{equation}

Since $g(u,u)=e\neq 0$, after contracting the equation with $g(u)$, we
obtain the condition 
\begin{equation}
ue-(\nabla _{u}g)(u,u)=0\text{ , \ \ \ \ or \ \ \ \ \ }ue=(\nabla _{u}g)(u,u)%
\text{ \ \ \ .}  \label{4.8}
\end{equation}

This condition determines how the length of the vector $u$ should change
with respect to the change of the metric $g$ along $u$ if $u$ should be an
auto-parallel vector field with $\nabla _{u}u=0$.

If we consider a Weyl's space as a model of space-time, this condition will
take the form 
\begin{equation}
ue=\frac{1}{n}\cdot Q_{u}\cdot e\text{ \ \ , or \ \ \ \ \ \ }u(\log e)=\frac{%
1}{n}\cdot Q_{u}\text{ \ \ ,\ \ }  \label{4.9}
\end{equation}
leading to the relation for $e$%
\begin{equation}
e=e_{0}\cdot \exp (\frac{1}{n}\cdot \int Q_{u}\cdot d\tau )\text{ \ \ , \ \
\ \ \ \ \ \ \ \ \ \ \ }e_{0}=\,\text{const.,}  \label{4.10}
\end{equation}
where $u=d/d\tau $ and $\tau =\tau (x^{k})$ is the canonical parameter of
the trajectory of the particle.

If $Q_{u}$ is constructed by the use of a dilaton field $\overline{\varphi }$
as $Q_{u}=-d\overline{\varphi }/d\tau $, then $e$ would change under the
condition 
\begin{equation}
e=e_{0}\cdot \exp (-\frac{1}{n}\cdot \overline{\varphi })\text{ .}
\label{4.11}
\end{equation}

The dilaton field $\overline{\varphi }$ could be represented by means of $e$
in the form 
\begin{equation}
\overline{\varphi }=-n\cdot \log (\frac{e}{e_{0}})\text{ .}  \label{4.12}
\end{equation}

Therefore, the dilaton field $\overline{\varphi }$ takes the role of a
length scaling factor for the velocity of a test particle.

(b) From the second equation (b) for $u$, it follows that a necessary and
sufficient condition for a free spinless test particle to move in a space
with affine connections and metrics on a trajectory, described by the
auto-parallel equation in its canonical form ($\nabla _{u}u=0$), is the
condition 
\begin{equation}
\lbrack u(\log \rho )]\cdot u=-\frac{1}{2}\cdot \overline{g}(\nabla _{u}g)(u)%
\text{ \ \ .}  \label{3.6}
\end{equation}

Since $g(u,u)=e\neq 0$, \ after contracting the last equation with $g(u)$ we
obtain the condition 
\begin{equation*}
u(\log \rho )\cdot e=-\frac{1}{2}\cdot \overline{g}(\nabla _{u}g)(u)[g(u)]=-%
\frac{1}{2}\cdot (\nabla _{u}g)(u,u)\text{\ \ ,}
\end{equation*}
or 
\begin{equation}
u(\log \rho )=-\frac{1}{2\cdot e}\cdot (\nabla _{u}g)(u,u)\text{ .}
\label{3.6a}
\end{equation}

For $u=d/d\tau $, it follows the equation 
\begin{equation*}
\frac{d}{d\tau }(\log \rho )=-\frac{1}{2\cdot e}\cdot (\nabla _{u}g)(u,u)%
\text{ ,}
\end{equation*}
with the solution for $\rho (x^{k}(\tau ))$%
\begin{equation*}
\rho =\rho _{0}\cdot \exp (-\frac{1}{2}\cdot \int \frac{1}{e}\cdot (\nabla
_{u}g)(u,u)\cdot d\tau )\text{ \ .}
\end{equation*}

The last condition is for the rest mass density $\rho $ which it has to obey
if the particle should move on an auto-parallel trajectory or if we observe
the motion of a particle as a free motion in the corresponding space
considered as a mathematical model of the space-time.

If we consider a Weyl's space as a model of the space-time, the condition (%
\ref{3.6}) will take the form 
\begin{equation}
\lbrack u(\log \rho )]\cdot u=-\frac{1}{2\cdot n}\cdot Q_{u}\cdot u\text{ ,}
\label{3.7}
\end{equation}
or the form 
\begin{equation*}
\lbrack u(\log \rho )+\frac{1}{2\cdot n}\cdot Q_{u}]\cdot u=0\text{ .}
\end{equation*}

Since $g(u,u)=e\neq 0$, \ after contracting the last equation with $g(u)$ we
obtain the condition 
\begin{equation}
u(\log \rho )+\frac{1}{2\cdot n}\cdot Q_{u}=0\text{ \ .}  \label{3.8}
\end{equation}

Therefore, the rest mass density $\rho $ should change on the auto-paralel
trajectory of the particle as 
\begin{equation}
\rho =\rho _{0}\cdot \exp \left[ -\frac{1}{2\cdot n}\cdot \int Q_{u}\cdot
d\tau \right] \text{ , \ \ \ \ \ \ \ \ }\rho _{0}=\text{const. }  \label{3.9}
\end{equation}

Furthermore, if $Q_{u}$ is constructed by the use of a dilaton field $%
\overline{\varphi }$ as $Q_{u}=-d\overline{\varphi }/d\tau $, then $\rho $
would change under the condition 
\begin{equation}
\rho =\rho _{0}\cdot \exp \left[ \frac{1}{2\cdot n}\cdot \int \frac{d%
\overline{\varphi }}{d\tau }\cdot d\tau \right] =\rho =\rho _{0}\cdot \exp (%
\frac{1}{2\cdot n}\cdot \overline{\varphi })\text{ .}  \label{3.10}
\end{equation}

The dilaton field $\overline{\varphi }$ could be represented by means of $%
\rho $ in the form 
\begin{equation}
\overline{\varphi }=2\cdot n\cdot (\log \frac{\rho }{\rho _{0}})\text{ .}
\label{3.11}
\end{equation}

Therefore, the dilaton field $\overline{\varphi }$ takes the role of a mass
density scaling factor for the rest mass density of a test particle. This is
another physical interpretation as the interpretation used by other authors
as mass field, pure geometric gauge field and etc.

Since $\overline{\varphi }=-n\cdot \log (e/e_{0})=2n\cdot \log (\rho /\rho
_{0})$, a relation between $e$ and $\rho $ follows in the form 
\begin{equation}
\rho ^{2}\cdot e=\rho _{0}^{2}\cdot e_{0}=\text{ const. }=l_{p}^{2}\text{ ,}
\label{4.13}
\end{equation}
which is exactly the condition (b) of the definition for a free moving
spinless test particle.

\section{Conclusion}

In the present paper the conditions are found under which a space with
affine connections and metrics and especially a Weyl's space admit
shear-free and expansion-free non-null vector fields. In a Weyl's space the
vector fields appear as conformal Killing vector fields. In such type of
spaces only the rotation velocity is not vanishing. This fact could be used
as a theoretical basis for models in contunuous media mechanics and in the
modern gravitational theories, where a rotation velocity could play an
important role. Further, necessary and sufficient conditions are found under
which a free spinless test particle could move in spaces with affine
connections and metrics on an auto-parallel curve. In Weyl's spaces with
Weyl's covector, constructed by the use of a dilaton field, the dilaton
field appears as a scaling factor for the rest mass density as well as for
the velocity of the test particle. The last fact leads to a new physical
interpretation of a dilaton field in classical field theories over spaces
with affine connections and metrics and especially over Weyl's spaces.

\end{document}